\documentclass{article}

\usepackage{arxiv}

\usepackage[utf8]{inputenc} 
\usepackage[T1]{fontenc}    
\usepackage{hyperref}       
\usepackage{url}            
\usepackage{booktabs}       
\usepackage{amsfonts}       
\usepackage{nicefrac}       
\usepackage{microtype}      
\usepackage{lipsum}
\usepackage{graphicx}
\graphicspath{ {./images/} }

\usepackage{algorithm}
\usepackage{algorithmic}
\usepackage{amsfonts}
\usepackage{amsmath}
\newtheorem{Definition}{Definition}
\newtheorem{lemma}{Lemma}
\usepackage{multirow}

\usepackage{newfloat}
\usepackage{listings}

\usepackage{graphicx}

\usepackage{hyperref}
\usepackage{balance}

\date{}

\begin{document}

\title{Joint Auction in the Online Advertising Market}

\author{
    Zhen Zhang\thanks{Both authors contribute equally to this research.} \\
    \texttt{zhangzhen2023@ruc.edu.cn} \\
    Gaoling School of  \\ Artificial Intelligence, \\ Renmin University of China \\
    Beijing, China 
    \And
    Weian Li\footnotemark[1] \\
    \texttt{weian.li@sdu.edu.cn} \\
    School of Software, \\ Shandong University \\
    Jinan, China 
    \And
    Yahui Lei \\
    \texttt{leiyahui@meituan.com}\\
    Meituan Inc. \\
    Beijing, China
    \And
    \\
    \textbf{Bingzhe Wang} \\
    \textbf{Zhicheng Zhang} \\
    \texttt{{wbz2022,mzhangzhicheng}@} \\
    ruc.edu.cn \\
    Gaoling School of \\ Artificial Intelligence, \\ Renmin University of China \\
    Beijing, China 
    \And
    \\
    \textbf{Qi Qi}\thanks{Corresponding author.} \\
    \texttt{qi.qi@ruc.edu.cn} \\ 
    Gaoling School of \\ Artificial Intelligence, \\ Renmin University of China \\
    Beijing, China 
    \And
    \\
    \textbf{Qiang Liu} \\
    \textbf{Xingxing Wang} \\
    \texttt{{liuqiang43,wangxingxing04}@} \\ 
    meituan.com\\
    Meituan Inc. \\
    Beijing, China 
}

\maketitle

\begin{abstract}
Online advertising is a primary source of income for e-commerce platforms. In the current advertising pattern, the oriented targets are the online store owners who are willing to pay extra fees to enhance the position of their stores. On the other hand, brand suppliers are also desirable to advertise their products in stores to boost brand sales. However, the currently used advertising mode cannot satisfy the demand of both stores and brand suppliers simultaneously. To address this, we innovatively propose a joint advertising model termed ``Joint Auction'', allowing brand suppliers and stores to collaboratively bid for advertising slots, catering to both their needs. However, conventional advertising auction mechanisms are not suitable for this novel scenario. In this paper, we propose JRegNet, a neural network architecture for the optimal joint auction design, to generate mechanisms that can
achieve the optimal revenue and guarantee (near-)dominant strategy incentive compatibility and individual rationality. Finally, multiple experiments are conducted on synthetic and real data to demonstrate that our proposed joint auction significantly improves platform’s revenue compared to the known baselines.
\end{abstract}


\section{Introduction}

Advertising is a primary source of income for internet companies such as Google, Amazon, and Facebook, bringing in significant amounts of revenue and supporting the development and prosperity of the internet. One widely-used method for assigning advertising positions is sponsored search auction, where advertisers submit bids to the platform, which then executes preset auction mechanisms to decide on advertising positions and charged fees. The monetization efficiency of advertising traffic directly affects the development of internet companies. How to improve the monetization efficiency of advertising has become a core research issue for enterprises. 

\begin{figure*}[h]
\centering
\includegraphics[width=0.85\textwidth]{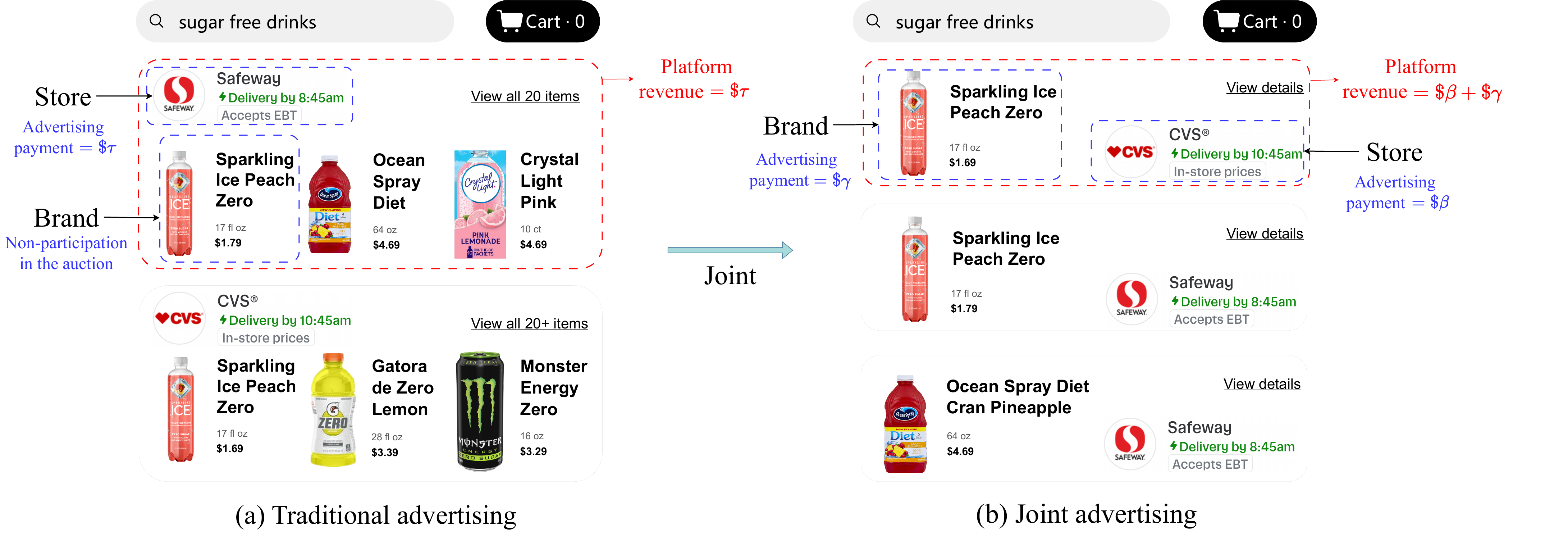}
\caption{\label{fig:1}An Example of the Traditional Advertising Model and Joint Advertising Model.}
\end{figure*}

In the context of e-commerce platform as an example, such as Instacart, when an inquiry arrives, the platform returns a list of stores that sell the related products (as shown in Figure \ref{fig:1}(a)). Stores can pay extra advertising fees to compete for higher positions in the list. 
From the perspective of brand suppliers, they also have a strong desire to enhance their exposure in order to drive up brand sales. Nevertheless, the current advertising model does not consider brand suppliers as advertisers, which causes that the platform may lose a portion of potential revenue from brand suppliers.

Motivated by this phenomenon and to further increase platform revenue, we set up a novel online advertisement model called “joint advertising” catering for the demand of brand suppliers, as shown in Figure \ref{fig:1}(b). In this joint advertising model, both stores and brands participate in the auction, a sponsored advertising item is contributed by a bundle consisting of a store and a brand supplier, and we name the auction as “joint auction”. In contrast to the traditional advertising auction, joint auction not only provides the chance for brand suppliers to boost brand sales but also enhance the revenue of platform since it now charges for both stores and brand suppliers. In addition, joint advertising also benefits for
users, because it directly presents the searched products rather than the related stores. Therefore, the joint advertising scenario essentially aligns with the vested interests of multiple stakeholders,
including platforms, brands, and users, which achieves a triple-win. Joint auction enriches the practicability of traditional auction theory, potentially exerting a significant impact in both academia and industry. 

However, from the perspective of mechanism design, the new model of joint auction also brings new technical challenges. For example, how to overcome the dependency of bundle bids and how to decide the payments of two parties within a bundle. It leads that most of the classic and common-used mechanisms may not apply for the joint advertising. Compared to the difficulties encountered in theory, in recent years, with the rapid development of machine learning, the concept of automated mechanism design has been proposed in computing the optimal mechanism \cite{dutting2019optimal,rahme2021permutation,duan2022context}. However, the current neural network architectures are not suitable for joint auctions, mainly due to that
they do not deal with the allocation of bundles. Therefore, how to deal with the problem of designing the optimal joint auctions is deserved to be investigated.

\subsection{Main Contributions}
The advertising system's most pivotal technology is its sales mechanism. The industry has seen significant maturity in the traditional advertising auction mechanism, with most studies emphasizing revenue enhancement through the introduction of innovative variations to existing frameworks. Our study uniquely presents a practical and revenue-boosting advertising sales model known as the ``joint auction''. Nonetheless, many standard and widely-used mechanisms may not be applicable to this novel joint advertising model. To identify the optimal mechanisms that are both dominant strategy incentive compatible (DSIC) and individually rational (IR), we introduce JRegNet, a neural network architecture to generate the optimal satisfactory mechanisms. Subsequently, we validate our proposed architecture through numerous experiments on both synthetic and real-world datasets. These experiments demonstrate the superior performance of the generated mechanisms compared to established baselines. In summary, our primary contributions are as follows:

\textbf{Joint auctions.} 
In this paper, we introduce a novel joint advertising model that satisfies the marketing requirements of both stores and brands simultaneously, leading to an increase in revenue. Within this framework, we present the innovative ``joint auction'' mechanism, where both stores and brand suppliers  submit bids and compete for advertising slots. A distinct feature of the joint auction is to display a bundled advertisement consisting of a store and a brand within a single advertising slot, rather than a single advertiser. However, not all stores and brands can form such a bundle, as it is based on an established sales relationship. To design an optimal joint auction, we frame it as a learning problem and employ automated mechanism design techniques to generate the optimal mechanism. 

\textbf{JRegNet.} 
To overcome the technical challenges in finding optimal joint auctions, we introduce an innovative neural network architecture called \textbf{J}oint \textbf{Reg}ret \textbf{Net}work (JRegNet). This architecture leverages the concept of regret to generate mechanisms that not only adhere to the IR and DSIC conditions but also maximize revenue. Importantly, JRegNet is capable of addressing the unique challenges posed by joint auctions, which are not effectively handled by current popular neural network architectures (e.g., \cite{feng2018deep,rahme2021permutation,duan2022context}):

(1) \textbf{Joint relationship graph.} Not all stores and brands can be combined into bundles, and the joint auction relationship can vary depending on different search queries within the same auction scenario. To tackle this issue, JRegNet incorporates the joint relationship graph of stores and brands as an input. This relationship graph plays a crucial role throughout all stages of JRegNet, influencing the formation of bundles.

(2) \textbf{Implementation of bundle constraints.} Joint auctions impose certain constraints on bundles, such as limiting each advertising slot to a single bundle and vice versa. To address these constraints, JRegNet initially computes an initial allocation probability matrix for the bundles. Subsequently, it employs two \textit{softmax} functions to ensure compliance with the bundle constraints.


(3) \textbf{Correlated bids of different bundles.} Since the advertising slot is ultimately allocated to a bundle, from the perspective of bundles, bids are dependent due to the potential presence of a single store or brand in multiple bundles. JRegNet deals with the problem of correlated bids by directly incorporating the independent bids of each store and brand at the input stage. This approach allows JRegNet to directly calculate the allocation probabilities of bundles based on these independent bids, rather than relying on the bids of bundles themselves within the neural network.

(4) \textbf{Calculation of separate payment.} Determining the payment for each bidder within a bundle can be a sophisticated problem, especially when the mechanism assigns a bundle to a specific slot. To solve this challenge, JRegNet 
introduces a parameter that scales the total expected value of each bidder based on the advertiser's allocation probability matrix (which is calculated by the bundle's allocation probability matrix) and defines the scaled value as the payment. This parameter is optimized through training, while ensuring the condition of IR for all bidders.

\subsection{Related Work}

Traditional auction mechanisms have found extensive application within the advertising auction domain. The classic Myerson auction \cite{myerson1981optimal} endeavors to maximize revenue in the single-parameter setting, which can apply to sponsored advertising directly.  
On the other hand, Vickrey-Clarke-Groves (VCG) mechanism \cite{vickrey1961counterspeculation, clarke1971multipart, groves1973incentives} is engineered to maximize social welfare of advertisers and does not depend on prior distributions, while keeping IR and DSIC. 
In recent two decades, generalized second price (GSP) auction \cite{varian2007position, edelman2007internet, caragiannis2011efficiency, gomes2009bayes, lucier2012revenue} is the most popular mechanism in industry, owing to its simplicity and comprehensibility. 
In pursuit of augmenting revenue of GSP, numerous researchers contribute to studying the variants of GSP. Lahaie and Pennock \cite{lahaie2007revenue} introduce  the concept of ``squashing'' to GSP, which assesses bidder rankings using a compression score during the allocation phase. Thompson and Leyton-Brown \cite{thompson2013revenue} integrate the idea of ``squashing'' and ``reserve price'' into GSP. Roberts et al. \cite{roberts2016ranking} propose a mechanism that ranks bidders based on the discrepancy between their bids and the reserve price. Charles et al. \cite{charles2016multi} conduct an exhaustive investigation into the impact of the squashing factor on revenue and CTR. 
However, as far as we know, except the VCG mechanism, all the other common mechanisms may not be applicable to the context of joint auctions (e.g., difficulty to define the equilibrium of GSP or dependent bids making Myerson auction invalid). In this paper, we design a novel mechanism for the joint auctions, while maintaining DSIC and IR, and yielding the optimal revenue. 

With the development of the field of machine learning, the approach of ``automated mechanism design'' (AMD) \cite{conitzer2002complexity, conitzer2004self, sandholm2015automated} is proposed for auction design, particularly for multi-item auctions. Different from the theoretical methods, AMD focuses more on using the techniques of machine learning to compute the approximate optimal auction mechanisms \cite{balcan2008reducing, lahaie2011kernel, dutting2015payment}. 
As a milestone work of AMD, D{\"u}tting et al. \cite{dutting2019optimal} model the problem of truthful auction design into a mathematical programming, transferring DSIC condition into the constraints on regrets and come up with the first neural network framework, RegretNet, which can achieve the optimal revenue. Later, there have been many works that extend RegretNet to handle different constraints and objectives, such as budget constraints \cite{feng2018deep}, fairness objective \cite{kuo2020proportionnet} and human preferences \cite{peri2021preferencenet}. 
Rahme et al. \cite{rahme2021permutation} propose a permutation-equivariant architecture, EquivariantNet, for designing the symmetric auctions. A transformer-based neural network architecture, CITransNet, catering for the contextual auctions is investigate by Duan et al. \cite{duan2022context}. Ivanov et al. \cite{ivanov2022optimal} modify the loss function and propose the RegretFormer architecture based on attention layers. Most existing results of AMD are under the assumptions that one item is allocated to at most one bidder and the bids on different items and from different bidders are independent, which are not suitable for joint auction, since one item is assigned to a bundle (consisting of two bidders) and the bids of different bundles are dependent. Additionally, in the same joint auction scenario, the joint auction relationship varies across different search queries, making it difficult for the existing AMD architecture to be applied to joint auction scenarios. Our proposed architecture, JRegNet, can overcome these difficulties and generate the near-optimal revenue in the joint auction.

\section{Joint Auction Design}
\label{sec:pre}
In this section, we first formally introduce the model of joint auction. Then, we transfer the optimal mechanism design problem into a learning problem. 
\subsection{Joint Auction}




In the context of joint auctions, whenever a user searches for a keyword, the platform returns an interface containing $K$ advertising slots, where each advertising slot displays the information of an advertising bundle consisting of a store and a brand. Denote by $\alpha_k$ the CTR of the $k$-th slot for any $k \in \{1, \ldots , K \}$. W.l.o.g., we assume that $1>\alpha_1 \ge \cdots \ge \alpha_K>0$. 

\begin{figure}[h]
\centering
\includegraphics[width=0.4\textwidth]{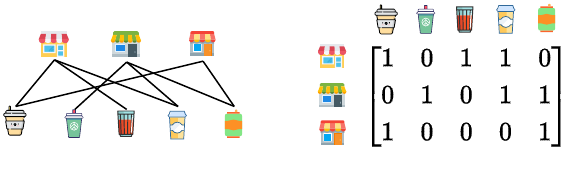}
\caption{\label{shopp}The bundle relationship between stores and brands. We use different beverage icons to represent different brands, and an edge linking a store and a brand means that there is a sales relationship between the two.}
\end{figure}

There are $m$ stores and $n$ brands participating in a joint auction. For store $i \in M=\{1, \ldots, m \}$ and brand $j \in N=\{1, \ldots, n \}$, we define an indicator, $\mathbf{1}_{ij}$,  to represent whether store $i$ and brand $j$ can form a bundle. Specifically, $\mathbf{1}_{ij} = 1$ means that  store $i$ and brand $j$ have a cooperative relationship. Moreover, we use a matrix to demonstrate the bundle relationship between all stores and all brands, where each entity is an indicator defined on a pair of store and brand (see Figure \ref{shopp} as an example). Within the same joint auction scenario, the joint auction relationship in the matrix varies across different search query samples. 


For each store $i$ (or brand $j$), the value for each click is denoted by $v_{i \cdot}$ (or $v_{\cdot j}$) which is private information of store $i$ (or brand $j$). Suppose $v_{i \cdot}$ (or $v_{\cdot j}$) is drawn from a commonly known distribution over the domain set $V_{i \cdot}$ (or $V_{\cdot j}$).  
The value profile can be expressed as $\boldsymbol{v} = (v_{1 \cdot}, \ldots, v_{m \cdot}, v_{\cdot 1}, \ldots, v_{\cdot n}) \in \mathbb{V}$, where $\mathbb{V} = V_{1 \cdot} \times \cdots \times V_{m \cdot} \times V_{\cdot 1} \times \cdots \times V_{\cdot n}$. Additionally, we use $\boldsymbol{v}_{-i \cdot}$ to stand for the value profile except for store $i$, 
i.e., $\boldsymbol{v}_{-i \cdot} = (v_{1 \cdot}, \ldots, v_{(i-1) \cdot}, v_{(i+1) \cdot}, \ldots, v_{\cdot n})$. 
The similar notations also can be defined on brand $j$. Since the store (or brand) may have a strategic bidding, we exploit $b_{i \cdot}$ (or $b_{\cdot j}$) to represent the bid price. Similarly, denote $\boldsymbol{b}$, $\boldsymbol{b}_{-i \cdot}$ and $\boldsymbol{b}_{\cdot -j}$ by the related bid profiles.   






In the setting of the joint auction, an auction mechanism, $\mathcal{M}(g,p)$, comprises two parts: the allocation rule $g$ and payment rule $p$. In detail, $g=\big((g_{i \cdot})_{i\in M}, (g_{\cdot j})_{j\in N} \big)$  where $g_{i \cdot}$ (or $g_{\cdot j}$) $: \mathbb{V} \to \mathbb{R}^{+} \cup \{0\}$ represents the expected CTR the store $i$ (or brand $j$) can derive. That is, $g_{i \cdot}(\boldsymbol{b}) = \sum_{k=1}^{K}{s_{i \cdot k}(\boldsymbol{b}) \alpha_k}$, where $s_{i \cdot k}(\boldsymbol{b})$ indicates the probability that store $i$ is assigned to $k$-th slot. The payment rule $p=\big((p_{i \cdot})_{i\in M}, (p_{\cdot j})_{j\in N} \big)$ means that how much the store $i$ or brand $j$ needs to charge. 
Different from the conventional advertising auctions, in a joint auction, each slot can be allocated to at most one bundle, and each bundle can get at most one slot. In other words, one store and one brand form a bundle to be displayed in a slot. 

Each store $i$ (or brand $j$) aims to maximize its own utility, which is defined in a quasi-linear form, i.e., $ u_{i \cdot}(v_{i \cdot} ; \boldsymbol{b}) = v_{i \cdot}(g_{i \cdot}(\boldsymbol{b})) - p_{i \cdot}(\boldsymbol{b}) $, for all $i\in M$ (the utility of brand $j$ has the similar form). In this paper, we focus on the mechanisms satisfying dominant strategy incentive compatibility and individual rationality. Roughly speaking, dominant strategy incentive compatibility guarantees that for any bidder, truthful bid can bring the maximum utility, i.e., 
\begin{Definition}[DSIC]
A joint auction is dominant strategy incentive compatible if any store $i$’s (or brand $j$'s) utility 
is maximized by reporting truthfully no matter what the others report. In other words, it holds that 
$$u_{i \cdot}(v_{i \cdot}; (v_{i \cdot}, \boldsymbol{b}_{-i \cdot})) \ge u_{i \cdot}(v_{i \cdot}; (b_{i \cdot}, \boldsymbol{b}_{-i \cdot})),$$ 
for all $ i \in M, v_{i \cdot} \in V_{i \cdot}, b_{i \cdot} \in V_{i \cdot}, \boldsymbol{b}_{-i \cdot} \in \mathbb{V}_{-i \cdot},$ and 
$$u_{\cdot j}(v_{\cdot j}; (v_{\cdot j}, \boldsymbol{b}_{-\cdot j})) \\ \ge u_{\cdot j}(v_{\cdot j}; (b_{\cdot j}, \boldsymbol{b}_{-\cdot j})),$$ for all $j \in N, v_{\cdot j} \in V_{\cdot j},  b_{\cdot j} \in V_{\cdot j},  \boldsymbol{b}_{-\cdot j} \in \mathbb{V}_{-\cdot j}$.
\end{Definition}
Individual rationality means that any participant of auction can get a non-negative utility, defined as follows,
\begin{Definition}[IR]
A joint auction is ex-post individually rational if any store $i$ (or brand $j$) receives a nonzero utility when participating truthfully, i.e., 
$$u_{i \cdot}(v_{i \cdot}; (v_{i \cdot}, \boldsymbol{b}_{-i \cdot})) \ge 0, \quad \forall i \in M, \forall v_{i \cdot} \in V_{i \cdot}, \forall \boldsymbol{b}_{-i \cdot} \in \mathbb{V}_{-i \cdot},$$ and $$u_{\cdot j}(v_{\cdot j}; (v_{\cdot j}, \boldsymbol{b}_{-\cdot j})) \ge 0, \quad \forall j \in N, \forall v_{\cdot j} \in V_{\cdot j}, \forall \boldsymbol{b}_{-\cdot j} \in \mathbb{V}_{-\cdot j}.$$
\end{Definition}

In a joint auction which is DSIC and IR, the expected revenue of platform is equal to the total payment of all stores and brands, defined as 
$$ rev := \mathbb{E}_{\boldsymbol{v} \sim F}\big[\sum_{i=1}^{m}{p_{i \cdot}(\boldsymbol{v})} + \sum_{j=1}^{n}{p_{\cdot j} (\boldsymbol{v})} \big],$$
where $F$ is the joint distribution of all stores' and brands' values. Optimal joint auction design aims to find an auction mechanism that maximizes the expected revenue while satisfying the DSIC and IR conditions.
\subsection{Joint Auction Design as a Learning Problem}










We formulate the problem of designing the optimal joint auction as a learning problem. First, to satisfy the DSIC condition, we introduce the concept of ex-post regret. Fixed others' bids, 
the ex-post regret for store $i$ (similar definition for brand $j$) is the highest increase on utility that she can obtain by misreporting, i.e.,
$$
rgt_{i \cdot}(\boldsymbol{v})=\mathbb{E}_{v \sim F}[\max_{v_{i \cdot}' \in V_{i \cdot}}{u_{i \cdot} (v_{i \cdot}; (v_{i \cdot}', \boldsymbol{v}_{-i \cdot}))} - u_{i \cdot} (v_{i \cdot};\boldsymbol{v})] \text{\hfill .}
$$
Thus, an auction mechanism satisfies the DSIC condition if and only if $rgt_{i \cdot}(\boldsymbol{v}) = 0$ and $rgt_{\cdot j}(\boldsymbol{v}) = 0$ are fulfilled, for all stores and brands. 
Moreover, we can formulate the problem of designing the optimal joint auction as a constrained optimization:
\begin{align}
\label{jj2}
\min_{(g,p) \in \mathcal{M}} \quad & -\mathbb{E}_{v \sim F}[\sum_{i=1}^{m}{p_{i \cdot}(\boldsymbol{v})} + \sum_{j=1}^{n}{p_j(\boldsymbol{v})}] \\
\text {s.t.} \quad & rgt_{i \cdot}(\boldsymbol{v}) = 0, \quad i = 1, \ldots, m,  \notag \\
& rgt_{\cdot j}(\boldsymbol{v}) = 0, \quad j = 1, \ldots, n \text{\hfill .} \notag
\end{align}
where $\mathcal{M}$ is the set of all joint auction mechanisms that satisfy IR. Due to the complex constraints, this optimization is generally intractable to solve \cite{conitzer2002complexity, conitzer2004self}. To solve this optimization problem, we parameterize the auction mechanisms as $\mathcal{M}^w (g^w, p^w) \subseteq \mathcal{M}(g, p)$ through parameter $w \in \mathbb{R }^{d}$ (where $d$ is the dimension of parameters $w$). Then, we move to  compute the mechanism $\mathcal{M}^w (g^w, p^w)$ which maximizes the expected revenue: $\mathbb{E}_{v \sim F}[\sum_{i=1}^{m}{p_{i \cdot}^w(\boldsymbol{v})} + \sum_{j=1}^{n}{p_{\cdot j}^w (\boldsymbol{v})}]$ and satisfies DSIC and IR, by optimizing the parameters $w$.

Given a sample $\mathcal{L}$, consisting of $L$ value profiles drawn from the joint distribution $F$, the empirical ex-post regret of store $i$ (similar notation for brand $j$) for the mechanism $\mathcal{M}^w (g^w, p^w)$ is estimated by:       
\begin{align}
\label{j2}
\widehat{r g t}_{i \cdot}(w)  = & \frac{1}{L} \sum_{l=1}^{L}[\max_{v_{i \cdot}^{\prime} \in V_{i \cdot}} u_{i \cdot}^w (v_{i \cdot}^{(l)}; (v_{i \cdot}^{\prime}, \boldsymbol{v}_{-i \cdot}^{(l)}))  \notag\\ 
& - u_{i \cdot}^w (v_{i \cdot}^{(l)}; \boldsymbol{v}^{(l)})] \text{\hfill .}
\end{align}
Rewrite the constrained optimization (\ref{jj2}) into:
\begin{align}
\label{j1}
\min_{w \in \mathbb{R}^{d_w}} \quad & -\frac{1}{L} \sum_{l=1}^{L}{[\sum_{i=1}^{m}{p_{i \cdot}^w(\boldsymbol{v}^{(l)})} + \sum_{j=1}^{n}{p_{\cdot j}^w(\boldsymbol{v}^{(l)})}]} \\
\text {s.t.} \quad & \widehat{rgt}_{i \cdot}(w) = 0, \quad i = 1, \ldots, m, \notag\\
& \widehat{rgt}_{\cdot j}(w) = 0, \quad j = 1, \ldots, n \text{\hfill .} \notag
\end{align}
In addition, through the network architecture, we ensure the IR condition and we will describe it in detail in Section \ref{sec:JA111}.

\section{JRegNet}
\label{sec:JA}

In this section, after modeling the optimal joint auction design as a learning problem, we propose a neural network architecture called \textbf{J}oint \textbf{Reg}ret \textbf{Net}work (JRegNet) for the optimal joint auction design.


\begin{figure*}[ht!]
\centering
\includegraphics[width=0.85\textwidth]{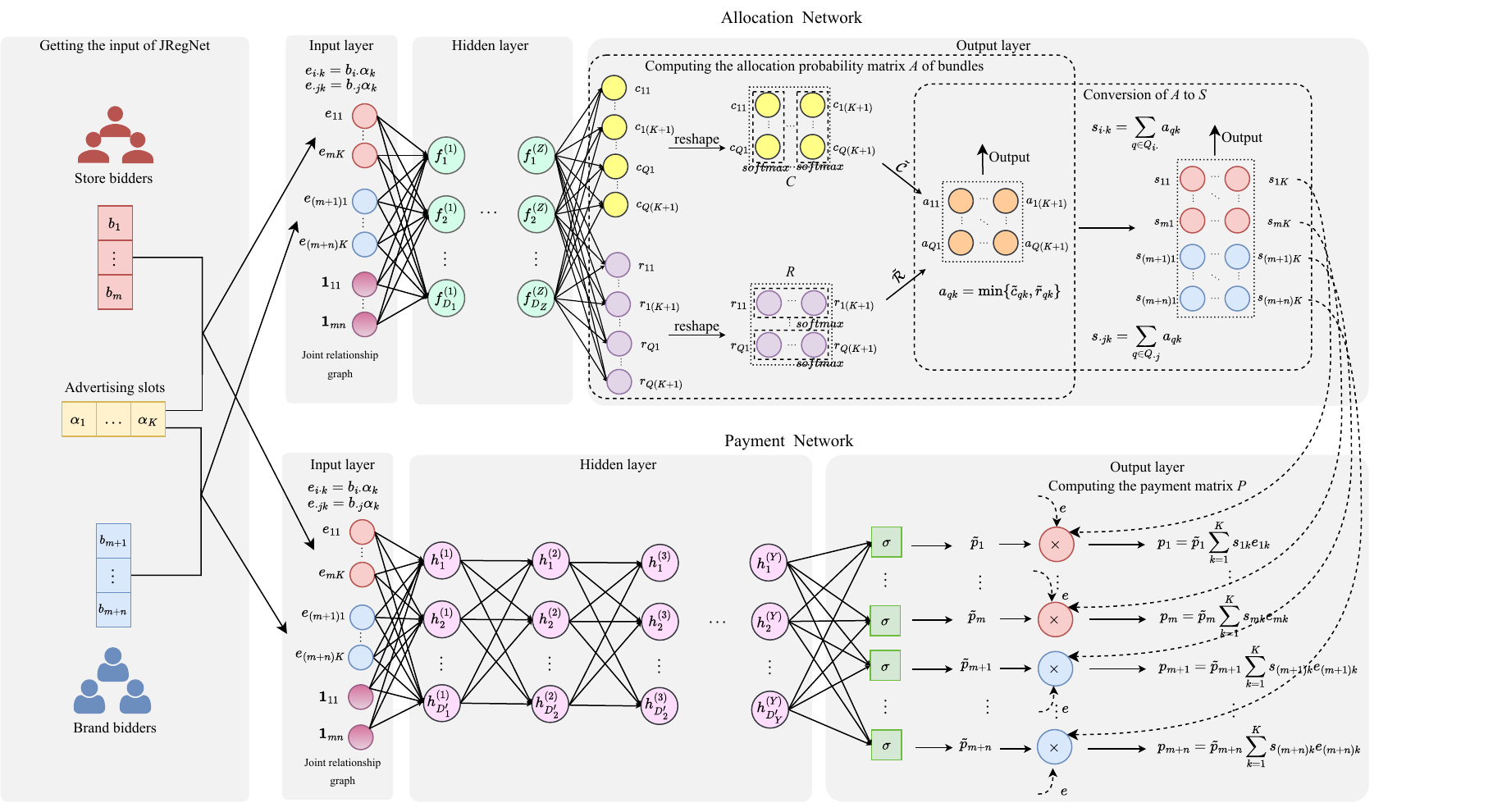}
\caption{\label{fig:3}The JRegNet architecture, including the allocation network part and the payment network part, for a setting of $m$ stores, $n$ brands, $K$ advertising slots, and $Q$ bundles of stores and brands.}
\end{figure*}

\subsection{The JRegNet Architecture}
\label{sec:JA111}

In this subsection, we introduce the network architecture of JRegNet designed for joint auction settings, as shown in Figure \ref{fig:3}. The JRegNet architecture consists of two parts: the allocation network encoding allocation rules and the payment network encoding payment rules. The outputs of both networks are used to calculate the utility of bidders, the revenue, the regret, and the loss function value of the entire network. Next, we introduce the main components of JRegNet in detail.


\textbf{Input and output of JRegNet.} In JRegNet, the neural network takes $\mathbf{1}_{ij}$, $e_{i \cdot k}$ and $e_{\cdot j k}$ as inputs, defined as:
\begin{equation}
\left\{\begin{array}{l}
e_{i \cdot k}=\alpha_{k} b_{i \cdot},\quad \forall i \in M,\forall k \in \{1,...,K \} \text{\hfill ,}\\
e_{\cdot j k}=\alpha_{k} b_{\cdot j},\quad \forall j \in N,\forall k \in \{1,...,K \} \text{\hfill ,}
\end{array}\right.
\label{me1}
\end{equation}
where $e_{i \cdot k}$ and $e_{\cdot j k}$ represent the expected bid for store $i$ and brand $j$ on slot $k$, respectively. $\{\mathbf{1}_{ij}\}_{i\in M, j\in N}$ represents the joint relationship matrix between stores and brands. Because the joint relationship varies across different search request samples, even under the same joint auction scenario (i.e., the same stores and brands), JRegNet takes $\{\mathbf{1}_{ij}\}_{i\in M, j\in N}$ as one of the inputs. In JRegNet, to adapt to the joint auction settings, the allocation network firstly   outputs the allocation probability matrix $A$ of bundles. Matrix $A$ contains the allocation probabilities for each bundle consisting of a store and a brand on each advertising slot. After obtaining the matrix $A$, the allocation network then  outputs the corresponding allocation probability matrix $S$ of bidders. Matrix $S$ contains the allocation probabilities for each store and brand on each advertising slot. Afterwards, the payment network   outputs the payment matrix $P$ of all bidders. Matrix $P$ contains the payments of each store bidder and brand bidder.


\textbf{Computing the allocation probability matrix $A$ of bundles.} To calculate the allocation probability matrix $S$ of bidders, we need to firstly calculate the allocation probability matrix $A$ of bundles, because we need to implement several constraints through $A$. And then, $S$ is computed based on $A$. Besides, matrix $A$ can be used to allocate advertising slots to bundles consisting of stores and brands. The input for this part is matrices $C$ and $R$ obtained through forward propagation, as shown by the yellow and purple neurons in Figure \ref{fig:3}. We use $Q$ to represent the total number of bundles, where bundle $q \in \{1, \ldots, Q \}$ corresponds to a row-by-row traversal of all the entries in the matrix consisting of $\mathbf{1}_{i j}$, with the bundle corresponding to the $q$-th entry with value $1$. The allocation probability of bundle $q$ on slot $k$ in the allocation probability matrix $A$ is defined as $a_{q k}$. In our model, the output matrix $A$ must satisfy two constraints: (a), each advertising slot can only be assigned to one bundle, i.e.,  $\sum_{q=1}^{Q}  {a}_{q k} \leq 1, \forall k \in \{1,...,K\}$, and (b), each bundle can be assigned to at most one advertising slot, i.e., 
$\sum_{k=1}^{K}  {a}_{q k} \leq 1, \forall q \in \{1,...,Q\}$. We enforce constraints (a) and (b) by designing the allocation probability matrix $A$ as a bi-stochastic matrix. To construct $A$ as a bi-stochastic matrix, some operations are required on $C$ and $R$. The matrix $C$ is column-normalized to matrix $\tilde{C}$ and the matrix $R$ is row-normalized to matrix $\tilde{R}$, where the normalization is achieved by the softmax function. Because in joint auction settings, a specific bundle consisting of a store and a brand may not be assigned to any advertising slot, we include a dummy neuron for each bundle in the normalization of the softmax function. The outputs of dummy neurons represent the probabilities of bundles not being assigned any advertising slot. The entries in matrices $\tilde{C}$ and $\tilde{R}$ are defined as $\tilde{c}_{q k}$ and $\tilde{r}_{q k}$, respectively. Then, we compute allocation probability of each bundle in the matrix $A$ by: $a_{q k}=\min\{\tilde{c}_{q k}, \tilde{r}_{q k}\}, \forall q \in \{1,...,Q \},\forall k \in \{1,...,K+1 \}$. According to Lemma \ref{lemma1}, the matrix $A$ constructed in this way is a bi-stochastic matrix. The specific formula for computing the allocation probability matrix $A$ is:
\begin{align*}
a_{q k}=\varphi_{q k}^{B S}\left(C, R\right)=\min \left\{\frac{e^{c_{q k}}}{\sum_{t=1}^{Q} e^{c_{t k}}}, \frac{e^{r_{q k}}}{\sum_{t=1}^{K+1} e^{r_{q t}}}\right\} \text{\hfill ,}
\end{align*}
where $\frac{e^{c_{q k}}}{\sum_{t=1}^{Q} e^{c_{t k}}}$ and $\frac{e^{r_{q k}}}{\sum_{t=1}^{K+1} e^{r_{q t}}}$ are the row normalization of $c_{q k}$ and the column normalization of $r_{q k}$, respectively, and the index $K+1$ corresponds to the situation where the bundle is not allocated any advertising slot. This part can be found in Figure \ref{fig:3}.

\begin{lemma}[\cite{dutting2019optimal}]
\label{lemma1}
The matrix $\varphi_{q k}^{B S}\left(c, r\right)$ is bi-stochastic $\forall c, r \in \mathbb{R}^{Q K}$. For any bi-stochastic matrix $a \in[0,1]^{Q K}$, $\exists c, r \in \mathbb{R}^{Q K}$ for which $a=\varphi_{q k}^{B S}\left(c, r\right)$.
\end{lemma}

\textbf{Conversion of allocation probability matrix $A$ to allocation probability matrix $S$.} In allocation probability matrix $S$ of bidders, the allocation probabilities of store $i$ and brand $j$ on slot $k$ are denoted as $s_{i \cdot k}$ and $s_{\cdot j k}$, respectively. After obtaining the allocation probability matrix $A$ of bundles, the allocation probabilities in $S$ are computed through:
\begin{align*}
\left\{\begin{array}{l}
s_{i \cdot k}=\sum_{q \in Q_{i \cdot}} a_{q k},\quad \forall k \in \{1,...,K \} \text{\hfill ,}\\
s_{\cdot j k}=\sum_{q \in Q_{\cdot j}} a_{q k}, \quad \forall k \in \{1,...,K \} \text{\hfill ,}
\end{array}\right.
\end{align*}
where $Q_{i \cdot}$ and $Q_{\cdot j}$ denote the sets of all bundles that include store $i$ and brand $j$, respectively, i.e., the allocation probability of store $i$ on slot $k$ is the sum of the allocation probabilities on slot $k$ for all bundles that include store $i$ and the calculation of the allocation probability of brand $j$ on slot $k$ is similar to this.

\textbf{Computing the payment matrix $P$.} After obtaining the allocation probability matrix $S$ of bidders, we propagate it to the payment network to calculate the payment matrix $P$, as shown in Figure \ref{fig:3}. In the matrix $P \in R_{\geq 0}^{I+J}$, the first $m$ rows and the last $n$ rows represent the payments for stores and brands, respectively. The payments of store $i$ and brand $j$ in $P$ are denoted as $p_{i \cdot}$ and $p_{\cdot j}$, respectively. The payments in $P$ are computed through:
\begin{align*}
\left\{\begin{array}{l}
p_{i \cdot}=\tilde{p}_{i \cdot}\left(\sum_{k=1}^{K} s_{i \cdot k} e_{i \cdot k}\right),\quad \forall i \in M \text{\hfill ,}\\
p_{\cdot j}=\tilde{p}_{\cdot j}\left(\sum_{k=1}^{K} s_{\cdot j k} e_{\cdot j k}\right), \quad \forall j \in N \text{\hfill .}
\end{array}\right.
\end{align*}
where $\tilde{p}_{i \cdot} \in[0,1]$ and $\tilde{p}_{\cdot j} \in[0,1]$ are the normalization factors calculated by the sigmoid function, as shown by the green squares in Figure \ref{fig:3}. Under the condition of satisfying DSIC, because of $\tilde{p}_{i \cdot} \in [0,1]$, and the utility of the store $u_{i \cdot}=\sum_{k=1}^{K} s_{i \cdot k} e_{i \cdot k}-p_{i \cdot}$, the utility of each store must be greater than or equal to zero, satisfying IR. The proof of IR for brand is similar.

\subsection{Training of JRegNet}

In this subsection, we introduce the training process of JRegNet, which includes converting the objective function, dividing training samples, finding the optimal misreports, and updating the model parameters and the Lagrange multipliers.

\textbf{Conversion from constrained optimization problems to unconstrained optimization problems.} In order to train JRegNet, we firstly need an objective function. The constrained optimization objective is shown in (\ref{j1}). We use the augmented Lagrangian method over the space of the neural network parameters $w$ to transform the constrained optimization problem into an unconstrained optimization problem, so that we can obtain the following objective function:
\begin{align*}
\mathcal{C}_{\rho}(w ; \boldsymbol{\lambda})= &-\frac{1}{L} \sum_{\ell=1}^{L}\left[\sum_{i=1}^{m} p_{i \cdot}^{w}\left(\boldsymbol{v}^{(\ell)}\right)+\sum_{j=1}^{n} p_{\cdot j}^{w}\left(\boldsymbol{v}^{(\ell)}\right)\right]  \\
& + \sum_{i=1}^{m}{\lambda_{i \cdot} \widehat{rgt}_{i \cdot}(w)} + \sum_{j=1}^{n}{\lambda_{\cdot j} \widehat{rgt}_{\cdot j}(w)} \\
& + \frac{\rho}{2}\sum_{i=1}^{m}{(\widehat{rgt}_{i \cdot}(w))^2} +\frac{\rho}{2} \sum_{j=1}^{n}{(\widehat{rgt}_{\cdot j}(w))^{2}} \text{\hfill ,}
\end{align*}
where $\boldsymbol{\lambda} \in \mathrm{R}^{n}$ is the Lagrange multiplier and $\rho>0$ is the penalty factor.

After obtaining the objective function required for training, we need to divide the samples in order to train JRegNet.

\textbf{Division of training samples.} The training sample $\mathcal{L}$ is randomly divided into minibatches of size $B$. We use $T$ to represent the total number of iterations. For each iteration $t \in \{1,...,T\}$, we sample a minibatch $\mathcal{L}_t$, which is denoted by $\mathcal{L}_t = \{v^{(1)}, \ldots, v^{(B)}\}$ and input it into JRegNet for training, until all the minibatches in this partition have been used. Afterwards, the training samples $\mathcal{L}$ are randomly partitioned again into new minibatches of size $B$. The above process is repeated until all iterations have been completed.

Afterwards, during training, to calculate the regret in $\mathcal{C}_{\rho}(w ; \boldsymbol{\lambda})$, we need to find the optimal misreports that maximize the regret.

\textbf{Finding the optimal misreports.} To calculate the optimal misreports, we use the method of gradient ascent. For each minibatch, the misreport $v_{i \cdot}^{\prime(\ell)}$ or $v_{\cdot j}^{\prime(\ell)}$ is calculated, for each store $i$ or brand $j$ and each valuation profile $\ell$, by taking multiple gradient ascents, and all the misreports are maintained to initialize the misreports in the next epoch. The formula for gradient ascent is as follows:

\begin{align*}
\left\{\begin{array}{l}
v_{i \cdot}^{\prime(\ell)}=v_{i \cdot}^{\prime(\ell)}+\left.\gamma \nabla_{v_{i \cdot}^{\prime}}\left[u_{i \cdot}^{w}\left(v_{i \cdot}^{(\ell)} ;\left(v_{i \cdot}^{\prime}, \boldsymbol{v}_{-i \cdot}^{(\ell)}\right)\right)\right]\right|_{v_{i \cdot}^{\prime}=v_{i \cdot}^{\prime(\ell)}} \text{\hfill ,}\\
v_{\cdot j}^{\prime(\ell)}=v_{\cdot j}^{\prime(\ell)}+\left.\gamma \nabla_{v_{\cdot j}^{\prime}}\left[u_{\cdot j}^{w}\left(v_{\cdot j}^{(\ell)} ;\left(v_{\cdot j}^{\prime}, \boldsymbol{v}_{-\cdot j}^{(\ell)}\right)\right)\right]\right|_{v_{\cdot j}^{\prime}=v_{\cdot j}^{\prime(\ell)}} \text{\hfill ,}
\end{array}\right.
\end{align*}
for some $\gamma>0$. 

\textbf{Updating the model parameters and the Lagrange multipliers.} Afterwards, we can calculate the value of ${C}_{\rho}(w ; \boldsymbol{\lambda})$ to perform backpropagation and update the neural network parameters $w$ to minimize ${C}_{\rho}(w ; \boldsymbol{\lambda})$. In addition, during the training process of the model, the update of the Lagrange multipliers is performed every fixed number of iterations:

\begin{align*}
\left\{\begin{array}{l}
\lambda_{i \cdot }^{t+1}=\lambda_{i \cdot }^{ {t }}+\rho_{t} \widetilde{r g t}_{i \cdot }\left(w^{ {t+1 }}\right) , \quad \forall i \in M\text{\hfill ,}\\
\lambda_{\cdot j }^{ {t+1 }}=\lambda_{\cdot j }^{ {t }}+\rho_{t} \widetilde{r g t}_{\cdot j}\left(w^{ {t+1 }}\right) , \quad \forall j \in N \text{\hfill ,}
\end{array}\right.
\end{align*}
where $t$ represents the number of iterations, and $\widetilde{r g t}_{i \cdot}(w)$ and $\widetilde{r g t}_{\cdot j}(w)$ represents the empirical regret calculated on the minibatch $S_t$ based on (\ref{j2}). The updates of the model parameters and the Lagrange multipliers are performed alternately. The specific and complete algorithm process of training JRegNet is shown in Algorithm $1$.

\begin{algorithm}[h!]
\caption{JRegNet Training} 
\begin{algorithmic}[1]
\STATE {\bf Input:} Minibatches 
$\mathcal{L}_1, . . . , \mathcal{L}_T$ of size B \\
\STATE {\bf Parameters:} 
$\forall t \in \{1,\ldots,T\}, \rho_t > 0, \gamma > 0, \eta > 0, \Gamma \in \mathbb{N}, T \in \mathbb{N},H \in \mathbb{N}
$ \\
\STATE {\bf Initialize:} 
$w^0 \in \mathbb{R}^{d}, \lambda^0 \in \mathbb{R}^{m+n}$
\FOR{$t=0$ to $T$} 
  \STATE Receive minibatch $\mathcal{L}_t = \{v^{(1)}, \ldots, v^{(B)}\}$ 
  \STATE Initialize misreport $v_{i \cdot}^{\prime(\ell)} \in V_{i \cdot},v_{\cdot j}^{\prime(\ell)} \in V_{\cdot j}, \forall \ell \in \{1,\ldots,B\}, i \in M,  j \in N$
  \FOR{$r=0$ to $\Gamma$} 
    \STATE $\forall \ell \in \{1,\ldots,B\}, i \in M,j \in N:$
     \STATE $\quad$ $
            v_{i \cdot}^{\prime(\ell)}=v_{i \cdot}^{\prime(\ell)}+\left.\gamma \nabla_{v_{i \cdot}^{\prime}}\left[u_{i \cdot}^{w}\left(v_{i \cdot}^{(\ell)} ;\left(v_{i \cdot}^{\prime}, \boldsymbol{v}_{-i \cdot}^{(\ell)}\right)\right)\right]\right|_{v_{i \cdot}^{\prime}=v_{i \cdot}^{\prime(\ell)}} $
      \STATE $\quad$ $
          v_{\cdot j}^{\prime(\ell)}=v_{\cdot j}^{\prime(\ell)}+\left.\gamma \nabla_{v_{\cdot j}^{\prime}}\left[u_{\cdot j}^{w}\left(v_{\cdot j}^{(\ell)} ;\left(v_{\cdot j}^{\prime}, \boldsymbol{v}_{-\cdot j}^{(\ell)}\right)\right)\right]\right|_{v_{\cdot j}^{\prime}=v_{\cdot j}^{\prime(\ell)}}  $
  \ENDFOR
  \STATE Compute regret gradient:$\forall \ell \in [B], i \in M, j \in N:$
  \STATE $\quad$ $g_{\ell, i \cdot}^{t}= \left. \nabla_{w}\left[u_{i \cdot}^{w}\left(v_{i \cdot}^{(\ell)} ;\left({v^{\prime}}_{i \cdot}^{(\ell)}, \boldsymbol{v}_{-i \cdot}^{(\ell)}\right)\right)-u_{i \cdot}^{w}\left(v_{i \cdot}^{(\ell)} ; \boldsymbol{v}^{(\ell)}\right)\right]\right|_{w=w^{t}}$
  \STATE $\quad$ $g_{\ell, \cdot j}^{t}= \left. \nabla_{w}\left[u_{\cdot j}^{w}\left(v_{\cdot j}^{(\ell)} ;\left({v^{\prime}}_{\cdot j}^{(\ell)}, \boldsymbol{v}_{-\cdot j}^{(\ell)}\right)\right)-u_{\cdot j}^{w}\left(v_{\cdot j}^{(\ell)} ; \boldsymbol{v}^{(\ell)}\right)\right]\right|_{w=w^{t}}$
  \STATE Compute Lagrangian gradient using Formula (\ref{eueu}) and update $w^t$:
  \STATE $\quad$ $w^{t+1} \leftarrow w^{t}- \eta \nabla_{w} \mathcal{C}_{\rho_{t}}\left(w^{t}, \lambda^{t}\right)$
  \STATE Update Lagrange multipliers once in $H$ iterations:
    \STATE $\quad$ {\bf if} {$t$ is a multiple of $H$} {\bf then} 
    \STATE $\quad$  $\quad$  $\lambda_{i \cdot}^{t+1} \leftarrow \lambda_{i \cdot}^{t}+\rho_{t} \widetilde{r g t}_{i \cdot}\left(w^{t+1}\right), \quad \forall i \in M$
     \STATE $\quad$  $\quad$  $\lambda_{\cdot j}^{t+1} \leftarrow \lambda_{\cdot j}^{t}+\rho_{t} \widetilde{r g t}_{\cdot j}\left(w^{t+1}\right), \quad \forall j \in N$
  \STATE $\quad$ {\bf else}
    \STATE $\quad$  $\quad$ $\boldsymbol{\lambda}^{t+1} \leftarrow \boldsymbol{\lambda}^t$
   \STATE $\quad$ {\bf end if}
\ENDFOR
\end{algorithmic}
\end{algorithm}

For fixed $\boldsymbol{\lambda}^t$, the gradient of $C_\rho$ w.r.t. $w$ can be written as:


\begin{align}
\label{eueu}
\quad \nabla_{w} \mathcal{C}_{\rho}\left(w ; \boldsymbol{\lambda}^{t}\right)= &-\frac{1}{B} \sum_{\ell=1}^{B} \left[\sum_{i=1}^{m} \nabla_{w} p_{i \cdot }^{w}\left(\boldsymbol{v}^{(\ell)}\right)+ \sum_{j=1}^{n} \nabla_{w} p_{\cdot j}^{w}\left(\boldsymbol{v}^{(\ell)}\right)\right] \notag\\
& + \sum_{\ell=1}^{B}\left[ \sum_{i=1}^{m}  \lambda_{i \cdot}^{t} g_{\ell, i \cdot}+\sum_{j=1}^{n} \lambda_{\cdot j}^{t} g_{\ell, \cdot j} \right] \notag\\
& +\rho \sum_{\ell=1}^{B} \left[\sum_{i=1}^{m}  \widetilde{r g t}_{i \cdot}(w) g_{\ell, i \cdot }+ \sum_{j=1}^{n} \widetilde{r g t}_{\cdot j}(w) g_{\ell, \cdot j}\right]  \text{\hfill ,}
\end{align}

where

\begin{align*}
\quad \quad
\left\{\begin{array}{l}
g_{\ell, i \cdot}=\nabla_{w}\left[\max _{v_{i \cdot}^{\prime} \in V_{i \cdot}} u_{i \cdot}^{w}\left(v_{i \cdot}^{(\ell)} ;\left(v_{i \cdot}^{\prime}, \boldsymbol{v}_{-i \cdot}^{(\ell)}\right)\right)-u_{i \cdot}^{w}\left(v_{i \cdot}^{(\ell)} ; \boldsymbol{v}^{(\ell)}\right)\right]  \text{\hfill ,}\\
g_{\ell, \cdot j}=\nabla_{w}\left[\max _{v_{\cdot j}^{\prime} \in V_{\cdot j}} u_{\cdot j}^{w}\left(v_{\cdot j}^{(\ell)} ;\left(v_{\cdot j}^{\prime}, \boldsymbol{v}_{-\cdot j}^{(\ell)}\right)\right)-u_{\cdot j}^{w}\left(v_{\cdot j}^{(\ell)} ; \boldsymbol{v}^{(\ell)}\right)\right] \text{\hfill .}
\end{array}\right.
\end{align*}

\section{Experiments}
\label{sec:exp}

In this section, empirical experiments are conducted in order to evaluate the joint advertising model and the effectiveness of JRegNet.  Our experiments are run on a compute cluster with NVIDIA Graphics Processing Unit (GPU) cores.

\textbf{Baseline methods.} We compare JRegNet with the following baselines:
\begin{itemize}
\item RegretNet \cite{dutting2019optimal}, a neural network architecture for near DSIC mechanism design in the traditional advertising auction settings (only include stroes) which can achieve the optimal revenue.
\item \textbf{I}ndependent \textbf{Reg}ret \textbf{Net}work (IRegNet), a neural network architecture for the optimal mechanism design in the advertising auction settings, where both stores and brands are independent candidates, competing for different slots. I.e., each slot either displays a store advertising or a brand advertising (this scenario does not exist in reality).

\item VCG \cite{vickrey1961counterspeculation, clarke1971multipart, groves1973incentives}, a classic mechanism satisfying DSIC and IR. In our experiments, we apply VCG mechanism to the joint advertising settings directly. 

\end{itemize}

Note that, while implementing the RegretNet, the bidders are only the stores. For other mechanisms, the bidders are both stores and brands. Other attributes for all mechanisms are the same. 

\textbf{Evaluation.} To assess the performance of each method, we utilize the average empirical ex-post regret of all bidders: $\widehat{r g t}:=\frac{1}{n+m} (\sum_{i=1}^{m} \widehat{r g t}_{i \cdot} + \sum_{j=1}^{n} \widehat{r g t}_{\cdot j})$, the empirical revenue: $rev := \frac{1}{L} \sum_{\ell=1}^{L}$ $\left[\sum_{i=1}^{m} p_{i \cdot}^{w}\left(v^{(\ell)}\right) + \sum_{j=1}^{n} p_{\cdot j}^{w}\left(v^{(\ell)}\right)\right]$ and the empirical social welfare:
$s w:=\frac{1}{L} \sum_{\ell=1}^{L}\left(\sum_{i=1}^{m} \sum_{k=1}^{K} s_{i \cdot k}^{(\ell)} \alpha_{k} v_{i \cdot}^{(\ell)} \\ + \sum_{j=1}^{n} \sum_{k=1}^{k} s_{\cdot j k}^{(\ell)} \alpha_{k} v_{\cdot j}^{(\ell)}\right)$. 

\begin{table*}[]
\begin{tabular}{cllllll}
\hline
\multirow{2}{*}{\textbf{Method}} & {\multirow{2}{*}{{\begin{tabular}[c]{@{}c@{}}\multicolumn{1}{c}{\textbf{A: 3 $\times$ 4 $\times$ 1}}\\  rev \quad \, sw \quad \, rgt\end{tabular}}}} & {\multirow{2}{*}{{\begin{tabular}[c]{@{}c@{}}\multicolumn{1}{c}{\textbf{B: 3 $\times$ 5 $\times$ 3}}\\  rev \quad \, sw \quad \, rgt\end{tabular}}}} & {\multirow{2}{*}{{\begin{tabular}[c]{@{}c@{}}\multicolumn{1}{c}{\textbf{C: 3 $\times$ 5 $\times$ 5}}\\  rev \quad \, sw \quad \, rgt\end{tabular}}}} & {\multirow{2}{*}{{\begin{tabular}[c]{@{}c@{}}\multicolumn{1}{c}{\textbf{D: 4 $\times$ 5 $\times$ 3}}\\  rev 
\quad \, sw \quad \, rgt\end{tabular}}}} \\
                                 & \multicolumn{1}{c}{}                                                                                      & \multicolumn{1}{c}{}                                                                                      & \multicolumn{1}{c}{}                                                                                      & \multicolumn{1}{c}{}                                                                                      & \multicolumn{1}{c}{}                                                                                      & \multicolumn{1}{c}{}                                             \\ \hline \\ [-10pt] \hline
RegretNet             &       0.305$^{\phantom{\dagger}}$  0.453 \, \textless{}0.001                                                                                                  &      0.298$^{\phantom{\dagger}}$   0.399        \, \textless{}0.001                                                                                             &                                      0.302$^{\phantom{\dagger}}$  0.403   \, \textless{}0.001                                                                 &     0.370$^{\phantom{\dagger}}$          0.488 \, \textless{}0.001                                                                                                      \\ \hline
IRegNet                        & 0.469$^{\phantom{\dagger}}$  0.540 \, \textless{}0.001                                                                                 & 0.536$^{\phantom{\dagger}}$    0.651 \,  \textless{}0.001                                                                                 & 0.551$^{\phantom{\dagger}}$   0.680 \, \textless{}0.001                                                                                  & 0.583$^{\phantom{\dagger}}$    0.690  \, \textless{}0.001                                                                                                                                                                 \\ \hline 
VCG                              &      0.433$^{\phantom{\dagger}}$  \textbf{0.908} \, \quad $-$                                                                                                     &                  0.445$^{\phantom{\dagger}}$  \textbf{1.154} \, \quad $-$                                                                                       &   0.400$^{\phantom{\dagger}}$   \textbf{1.256} \, \quad $-$                                                                                                      &     0.602$^{\phantom{\dagger}}$   \textbf{1.232}  \, \quad $-$                                                                                                                                                                                                                                                                                                                                                   \\ \hline
JRegNet                          & \textbf{0.509$^{\dagger}$} 0.741 \,  \textless{}0.001                                                                                & \textbf{0.641$^{\dagger}$}  0.955   \,  \textless{}0.001                                                                                 & \textbf{0.660$^{\dagger}$}  1.007 \, \textless{}0.001                                                                                 & \textbf{0.676$^{\dagger}$}     1.005  \,  \textless{}0.001                                                                                                                                                        \\ \hline
\end{tabular}
\caption{The results of experiments for Settings A to D. ``${\dagger}$'' indicates that the revenue has a significant improvement over other methods in paired t-test at $p < 0.05$ level.
}\label{tbl:1}
\end{table*}

\begin{table*}[]
\centering
\begin{tabular}{lclll}
\hline
\multirow{2}{*}{\textbf{Method}} & \multicolumn{1}{c}{\textbf{Uniform}} & \multicolumn{1}{c}{\textbf{Normal}} & \multicolumn{1}{c}{\textbf{Lognormal}} \\
\multicolumn{1}{c}{}                                                                           & \multicolumn{1}{l}{rev \quad \, sw \quad \, rgt}                              & rev \quad \, sw \quad \, rgt                             & rev \quad \, sw \quad \, rgt                                \\ \hline \\ [-10pt] \hline

                                                                  RegretNet           &   \multicolumn{1}{l}{0.298  \, 0.399 \, \textless{}0.001}                                     &  0.307 \, 0.412 \, \textless{}0.001                                 &  0.276  \, 0.390 \, \textless{}0.001
\\

IRegNet                        & \multicolumn{1}{l}{0.536  \,  0.651   \, \textless{}0.001}               & 0.496  \,  0.583 \, \textless{}0.001             & 0.501  \, 0.626  \, \textless{}0.001                           \\
                                                                  VCG                              & \multicolumn{1}{l}{0.445 \, \textbf{1.154}  \, \quad $-$}                                  &   0.492    \, \textbf{1.045}  \, \quad $-$                                &  0.433  \, \textbf{1.099}  \, \quad $-$                                                                         \\
                                                                  JRegNet                          & \textbf{0.641$^{\dagger}$}  0.955  \, \textless{}0.001               & \textbf{0.674$^{\dagger}$}   0.949  \, \textless{}0.001             & \textbf{0.612$^{\dagger}$}   0.945  \,   \textless{}0.001                                    \\ \hline
\end{tabular}
\caption{The results of experiments for different value distributions. The setting is $3$ stores and $5$ brands with $3$ advertising slots. ``${\dagger}$'' indicates that the revenue has a significant improvement over other methods in paired t-test at $p < 0.05$ level.} 
\label{tbl:2}
\end{table*}

\begin{table*}[]
\centering
\begin{tabular}{cllllll}
\hline
\multirow{2}{*}{\textbf{Method}} & {\multirow{2}{*}{{\begin{tabular}[c]{@{}c@{}}\multicolumn{1}{c}{\textbf{E1}}\\  rev \quad \quad sw \quad \quad ru\end{tabular}}}} & {\multirow{2}{*}{{\begin{tabular}[c]{@{}c@{}}\multicolumn{1}{c}{\textbf{E2}}\\  rev \quad \quad sw \quad \quad ru\end{tabular}}}} & {\multirow{2}{*}{{\begin{tabular}[c]{@{}c@{}}\multicolumn{1}{c}{\textbf{E3}} \\  rev 
\quad \quad sw \quad \quad ru\end{tabular}}}} \\
                                 & \multicolumn{1}{c}{}                                                                                      & \multicolumn{1}{c}{}                                                                                      & \multicolumn{1}{c}{}                                                                                      & \multicolumn{1}{c}{}                                                                                      & \multicolumn{1}{c}{}                                                                                      & \multicolumn{1}{c}{}                                             \\ \hline \\ [-10pt] \hline

VCG                              &   28.641 \, \textbf{51.172}  \quad  $-$                                                                                                                                                                     &   23.931 \,  \textbf{59.340}  \quad $-$                                                                                                      &     28.262  \, \textbf{58.257}   \quad $-$                                                                                                                                                                                                            \\ \hline                                                                                                                            
        \multirow{2}{*}{JRegNet}                     &      40.666$^{\dagger}$  48.510  \,  0.020                                                                                                                                                                                                      &   45.874$^{\dagger}$   56.193  \,  0.015                                                                                                      &     45.920$^{\dagger}$   55.502  \,  0.015

\\ 

                       & \textbf{43.329$^{\dagger}$}  49.066 \,  0.445                                                                                                                                                                & \textbf{49.005$^{\dagger}$}  57.070 \, 0.402                                                                                 & \textbf{48.867$^{\dagger}$}     56.421  \,   0.868                                                                                                                                        \\ \hline

GSP                              &   40.764 \, \textbf{51.172}  \,  \textbf{1.117}                                                                                                                                                                     &   44.206 \,  \textbf{59.340}  \, \textbf{1.083}                                                                                                      &     44.765  \, \textbf{58.257}   \, \textbf{1.039}

\\ \hline

\end{tabular}
\caption{The results of experiments for real data on the test set. In experiments E1 to E3, we train JRegNet using the data from Figures \ref{fig:9999999}(a) to \ref{fig:9999999}(c) respectively, and test JRegNet and the baseline methods using the corresponding data in Figure \ref{fig:888}. There are two groups of experimental data in JRegNet section, where the first row  with a very small $ru$ is compared with VCG, and the second row is used for comparison with GSP. ``${\dagger}$'' indicates that the revenue has a significant improvement over other methods in paired t-test at $p < 0.05$ level. }\label{tbl:999}
\end{table*}

\subsection{Synthetic Data}
\textbf{Implementation details.} 
We create training sets and the test sets for each experiment. In the training sets, the size $B$ of minibatches is $128$ and the numbers of minibatches and iterations are $5000$ and $200000$, respectively. In the test sets, the size of minibatches is $128$ and the number of minibatches is $100$. Thus, the size of the training sample $\mathcal{L}$ is $640000$ and the size of the testing sample is $12800$. 

In all synthetic data experiments, the joint relationship matrix $\{\mathbf{1}_{ij}\}_{i\in M, j\in N}$ between stores and brands is randomly generated for each search request sample. Furthermore, within a search request sample, if a store (or brand) has no joint relationship with any brand (or store), we default its bid to be zero, since such stores (or brands) cannot form a bundle to be displayed. 

In the testing of JRegNet, for each bidder, we perform gradient ascent on the $100$ different initial misreports for $2000$ iterations, to obtain $100$ empirical regrets. Then, we take the maximum value of these $100$ empirical regrets as the empirical regret of this bidder. All experimental results of RegretNet, IRegNet and JRegNet are obtained by taking the average of $3$ different runs on the testing samples.

\textbf{The comparison between JRegNet and the baselines.} To demonstrate the superiority of the joint advertising model and the effectiveness of JRegNet, we conduct comparative experiments between mechanism generated by JRegNet and baselines in the following settings:

\begin{enumerate}
\item[(A)] $3$ stores, $4$ brands and $1$ advertising slot with CTR $\boldsymbol{\alpha}=(0.7)$. The value of each store and brand is independently drawn from $U[0, 1]$.

\item[(B)] $3$ stores, $5$ brands and $3$ advertising slots with CTRs $\boldsymbol{\alpha}=(0.5, 0.3, 0.15)$. The value of each store and brand is independently drawn from $U[0, 1]$.

\item[(C)] $3$ stores, $5$ brands and $5$ advertising slots with CTRs $\boldsymbol{\alpha}=(0.5, 0.3, 0.15, 0.1, 0.03)$. The value of each store and brand is independently drawn from $U[0, 1]$.

\item[(D)] $4$ stores, $5$ brands and $3$ advertising slots with CTRs $\boldsymbol{\alpha}=(0.5, 0.3, 0.15)$. The value of each store and brand is independently drawn from $U[0, 1]$.
\end{enumerate}

We show the results of experiments under Settings A to D in Table \ref{tbl:1}. First, regarding to all mechanisms generated by the methods of automated mechanism design, the mechanisms generated by JRegNet achieve a significantly higher revenue, compared to the mechanisms generated by RegretNet and IRegNet, when the regret is less than 0.001. This demonstrates that the joint advertising model can derive more revenue than the traditional model. Additionally, as shown in Table \ref{tbl:1}, if we focus on the joint advertising settings, we can find that JRegNet can also obtain a significantly higher revenue than VCG, despite experiencing a measure of social welfare loss. It implies, for revenue, the good performance of mechanisms generated by JRegNet and the effectiveness of JRegNet.

To further evaluate the joint advertising model and the performance of JRegNet, we execute the experiments under the setting with different distributions, CTRs and numbers of advertisement slots.

\textbf{Different value distributions}. To verify the performance and stability of our proposed mechanism under different value distributions, we repeat experiments in three different value distributions on setting B. The value of each store and brand is independently drawn from three different distributions: Uniform distribution, $U[0, 1]$; Normal distribution $N(0.5,0.0256)$ and $v \in [0,1]$; Lognormal distribution $LN(0.1,1.44)$ and $v \in [0,1]$.

It can be seen from the experimental results in Table \ref{tbl:2} that for three different value distributions, JRegNet achieves the significantly highest revenue, compared to RegretNet, IRegNet and VCG mechanism. This demonstrates the stability and robustness of our mechanisms, when facing the different value distributions. 

\textbf{Different CTRs of slots}. Based on the Setting B, we adjust the values of CTRs and repeatedly run all mechanisms in the following three settings: 
B1. CTRs, $\boldsymbol{\alpha}=(0.4, 0.3, 0.15)$; B2. CTRs, $\boldsymbol{\alpha}=(0.5, 0.4, 0.15)$;  B3. CTRs, $\boldsymbol{\alpha}=(0.5, 0.4, 0.25)$. The experimental results are shown in Figure \ref{fig:6}(a) of Appendix \ref{ap3}.

From Figure \ref{fig:6} (a), it can be seen that as the CTR increases, the JRegNet's revenue also increases, and the impact of changing the CTR on the first slot is greater than the impact of changing the CTR on the last slot. For all settings in Figure \ref{fig:6} (a), the mechanisms generated by JRegNet all realize the highest revenue. This proves that even for different CTRs, JRegNet can still generate mechanisms that perform well in the joint advertising settings. 

\textbf{Different numbers of advertisement slots}. Based on the Setting D, we increase or decrease the number of advertisement slots to evaluate the performance of our mechanisms. The adjusted settings are described as follows: Setting D1. two advertising slots with CTRs $\boldsymbol{\alpha}=(0.5, 0.3)$; Setting D2. four advertising slots with CTRs $\boldsymbol{\alpha}=(0.5, 0.3, 0.15, 0.1)$; Setting D3. five advertising slots with CTRs $\boldsymbol{\alpha}=(0.5, 0.3, 0.15, 0.1, 0.03)$. 
The curves of revenue as the number of advertising slots changing are shown in Figure \ref{fig:6} (b) of Appendix \ref{ap3}. 

From Figure \ref{fig:6} (b), the mechanisms generated by JRegNet perform best among all mechanisms, which demonstrates that our mechanism can be suitable for the settings with different numbers of slots. In addition, the revenue derived by our mechanism strictly increases, with the increase on the number of slots. This phenomenon is not appeared on all mechanisms we execute. 

\subsection{Real-World Dataset}

We train and evaluate our model using real online auction log data from Meituan, an e-commerce platform. Each user search query corresponds to one advertising auction. In the real scenario, after the recall, rough ranking, and fine ranking stages, the advertising system selects about $10$ bundles as auction candidates and picks up at most $10$ stores and at most $10$ brands from candidates. Then, platform runs an auction mechanism to allocate at most $5$ slots. This is the real scenario currently being tested by the online e-commerce platform. Therefore, we focus on the setting with $10$ stores, $10$ brands and $5$ slots. Due to the variable number of advertisers in the original data, we trim or pad some data records to ensure that we have $10$ stores and $10$ brands in each auction sample. The padding operation is accomplished by adding advertisers whose value per click is $0$. We use $6912$ real samples for testing. The log data features we use primarily include the bid and id of each store and brand, the joint relationship between stores and brands, and the predicted CTR of each advertising slot.

In addition, using regret alone to measure DSIC is no longer applicable due to the different value function distributions between real data and simulated data. For the real joint auction data and our simulated data, there is a significant difference in the magnitude of bidders' utilities. If we simply use a fixed regret threshold, it will not accurately measure the degree of DSIC. Therefore, we use $r u:=\frac{1}{L} \sum_{\ell=1}^{L}\left[\left(\sum_{i=1}^{m} r g t_{i \cdot}^{(\ell)}+\sum_{j=1}^{n} r g t_{\cdot j}^{(\ell)}\right) /\left(\sum_{i=1}^{m} \mu_{i \cdot}^{(\ell)}+ \sum_{j=1}^{n} \mu_{\cdot j}^{(\ell)}\right)\right]$ $\,$ to assess DSIC where $r g t_{i \cdot}  :=\max_{v_{i \cdot}^{\prime} \in V_{i \cdot}} u_{i \cdot} (v_{i \cdot}; (v_{i \cdot}^{\prime}, \boldsymbol{v}_{-i \cdot})) - u_{i \cdot} (v_{i \cdot}; \boldsymbol{v})$ (similar definition for brand $j$). The meaning of $ru$ is the ratio of utility growth obtained from misreporting to the utility from reporting truthfully. For GSP, we employ an enumeration method \cite{liao2022nma} during the testing phase to get the misreport, where the misreport is equal to $\beta v$, with $\beta \in \{0, 0.1, 0.2, 0.3, \ldots, 1.9\}$. We use the misreport that can obtain the maximum utility from this set to calculate the GSP's regret.

We train JRegNet with real data from three distinct time periods, and for each 
time period, we test them using data from three different time points respectively. 
In Figures \ref{fig:9999999} and \ref{fig:888}  of Appendix \ref{ap666}, we show the detailed number of real search query samples used for training and testing JRegNet.
In Table \ref{tbl:999}, we present the performance of JRegNet, VCG, and GSP on the test set across all real data experiments, where GSP is the mechanism initially used online by Meituan in the joint auction scenario.

From Table \ref{tbl:999}, it can be seen that in all real data experiments, 
JRegNet can achieve significantly higher revenue than VCG at a very small $ru$ and JRegNet can also achieve significantly higher revenue than GSP when their $ru$ are close, and all the paired t-test is at the level of $p < 0.05$. These real data experiments demonstrate the effectiveness of JRegNet in real joint auction scenarios. Furthermore, it should be noted that GSP does not satisfy DSIC, which leads to increased complexity in advertisers' bidding strategies and difficulty to predict the equilibrium of GSP in joint auction scenario. 
In the experiment, the revenue of the GSP is calculated under the assumption that advertisers bid truthfully. However, in the equilibrium state, advertisers' bids are lower than their true values. Therefore, the actual revenue of GSP is lower than what is presented in the table. But since the equilibrium of GSP in the joint auction scenario is difficult to calculate, we choose to use an upper bound.

JRegNet can work in real-world scenarios. If we use a server with $1$ GPU and $32$ CPUs to run this real setting, it will take about $2$ to $4$ hours to run $30000$ iterations of training. However, the training of the JRegNet architecture can be performed offline. Therefore, although offline training may be relatively slow, invoking the trained JRegNet neural network model for forward propagation to obtain allocation and payment matrices is very fast (around $1$ to $3$ milliseconds per search query sample), which does not affect online deployment and decision-making.

\section{Conclusions}
\label{sec:conc}

In this paper, we propose a novel joint advertising model where stores and brands participate in a joint auction to jointly bid for advertising slots. To design the optimal joint auction mechanisms satisfying DSIC and IR, we propose JRegNet, a neural network architecture for generating the optimal joint auctions. We also conduct multiple experiments to demonstrate 
the superiority of mechanisms generated by JRegNet, 
compared to the baselines. 

In the future, an interesting direction is to apply joint auction to other scenarios such as live streaming sales, different brand collaborations, and so on.

\section*{Acknowledgments}

This work was supported by the Fundamental Research Funds for the Central Universities, and the Research Funds of Renmin University of China (No. 22XNKJ07), Meituan Inc. Fund, and Major Innovation \& Planning Interdisciplinary Platform for the “Double-First Class” Initiative, Renmin University of China.
The work was partially done at Engineering Research Center of Next-Generation Intelligent Search and Recommendation, Ministry of Education, Beijing Key Laboratory of Big Data Management and Analysis Methods, Gaoling School of Artificial Intelligence, Renmin University of China. We would like to thank Wanzhi Zhang for helpful comments and discussions.

\newpage
\bibliographystyle{unsrt}   
\bibliography{main}

\clearpage
\appendix

\section{Additional Experimental Information}
\label{ap222222}

\subsection{Detailed Experiment Results for Different CTRs and Different Numbers of Advertisement Slots.}
\label{ap3}

\begin{figure}[H]
  \centering
  \includegraphics[width=0.4\textwidth]{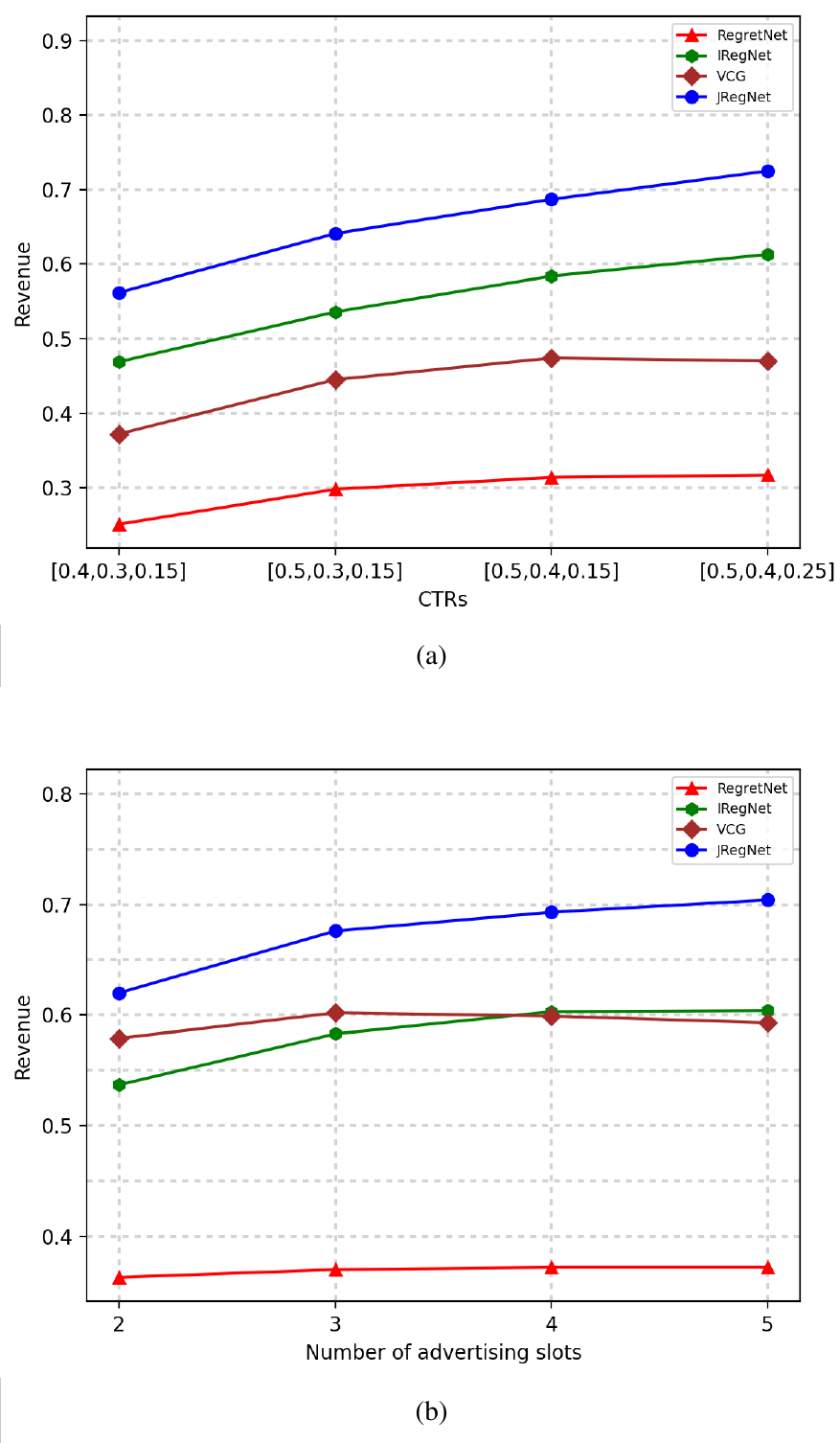}
  \caption{The results of experiments on the settings of different CTRs (a) and different numbers of advertisement slots (b). 
  In all results, the regrets are less than 0.001.} 
  \label{fig:6}
\end{figure}

The results of experiments on the settings of different CTRs and different numbers of advertisement slots are shown in Figure \ref{fig:6}. The detailed information about the experiment results is 
shown in Table \ref{tbl:666666} and Table \ref{tbl:333333}, respectively.

\subsection{Real-World Data Experiment}
\label{ap666}

\begin{figure}[H]
  \centering
  \includegraphics[width=0.4\textwidth]{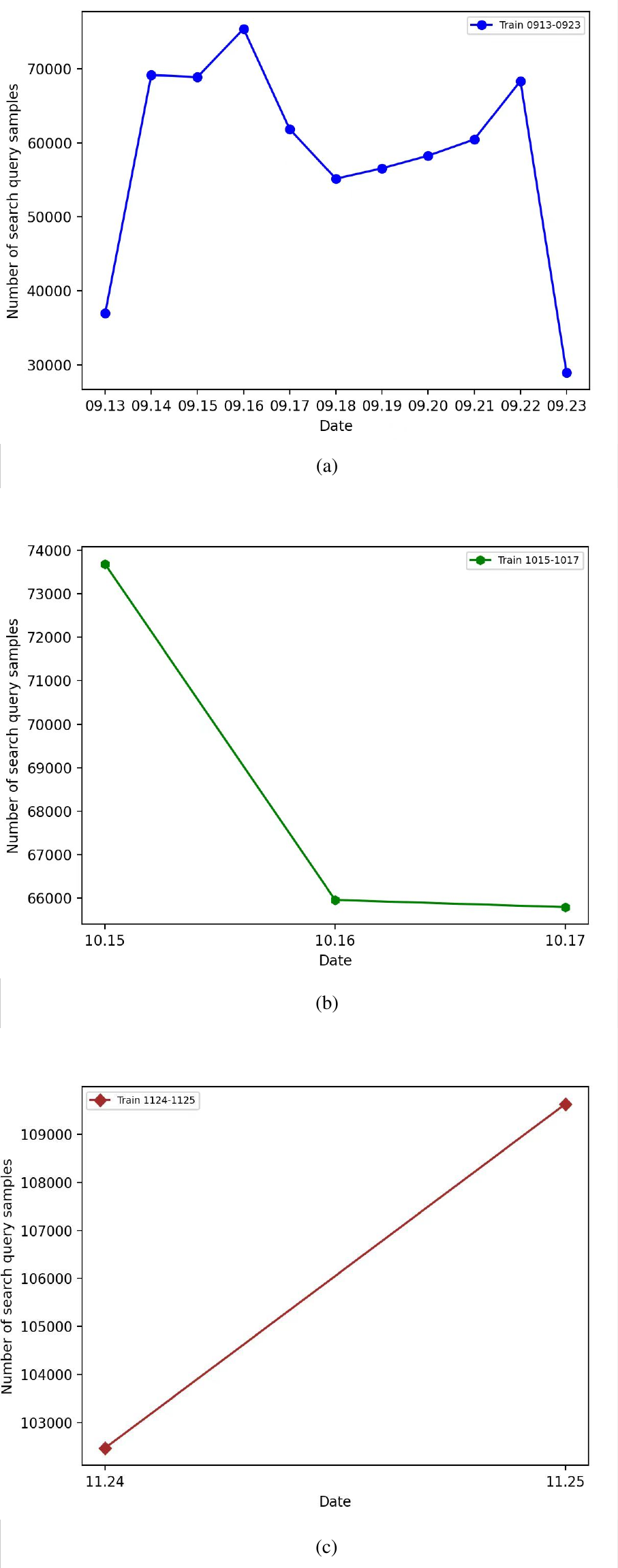}
  \caption{The number of real search query samples used for training JRegNet. The horizontal axis is the generation time of the search query samples. In this figure, all search query samples are obtained from logs of the year 2023. Within settings E1 to E3, we use data from (a) to (c) to train JRegNet, respectively.} 
  \label{fig:9999999}
\end{figure}

\begin{figure*}
  \centering
  \includegraphics[width=0.5\textwidth]{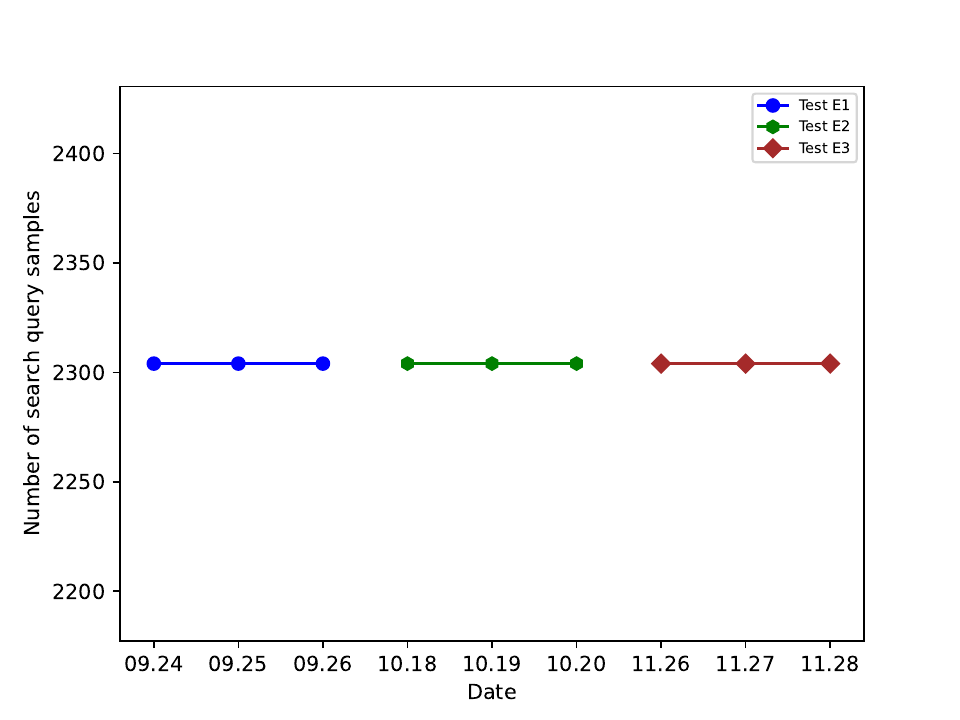}
  \caption{The number of real search query samples used for testing JRegNet and baseline methods. In this figure, all search query samples are obtained from logs of the year 2023. For settings E1 to E3, we take three days of real data for testing, respectively.} 
  \label{fig:888}
\end{figure*}

The number of real search query samples used for training JRegNet is shown in Figure \ref{fig:9999999}, and the number used for testing is shown in Figure \ref{fig:888}.

\begin{table*}[]
\begin{tabular}{cllllll}
\hline
\multirow{2}{*}{\textbf{Method}} & {\multirow{2}{*}{{\begin{tabular}[c]{@{}c@{}}\multicolumn{1}{c}{\textbf{B1: 3 $\times$ 5 $\times$ 3}}\\  rev \quad \, sw \quad \, rgt\end{tabular}}}} & {\multirow{2}{*}{{\begin{tabular}[c]{@{}c@{}}\multicolumn{1}{c}{\textbf{B: 3 $\times$ 5 $\times$ 3}}\\  rev \quad \, sw \quad \, rgt\end{tabular}}}} & {\multirow{2}{*}{{\begin{tabular}[c]{@{}c@{}}\multicolumn{1}{c}{\textbf{B2: 3 $\times$ 5 $\times$ 3}}\\  rev \quad \, sw \quad \, rgt\end{tabular}}}} & {\multirow{2}{*}{{\begin{tabular}[c]{@{}c@{}}\multicolumn{1}{c}{\textbf{B3: 3 $\times$ 5 $\times$ 3}}\\  rev \quad \, sw \quad \, rgt\end{tabular}}}} \\
                                 & \multicolumn{1}{c}{}                                                                                      & \multicolumn{1}{c}{}                                                                                      & \multicolumn{1}{c}{}                                                                                      & \multicolumn{1}{c}{}                                                                                      & \multicolumn{1}{c}{}                                                                                      & \multicolumn{1}{c}{}                                                                                \\ \hline \\ [-10pt] \hline
RegretNet           &       0.251   \, 0.340  \,  \textless{}0.001                                                                                                   &      0.298  \, 0.399   \, \textless{}0.001                                                                                              &                                      0.314         \,  0.428  \, \textless{}0.001                                                                  &    0.317    \, 0.433  \, \textless{}0.001                                                 \\  \hline
IRegNet                        & 0.469  \, 0.572  \, \textless{}0.001                                                                                 & 0.536  \,  0.651  \,  \textless{}0.001                                                                                 & 0.584   \, 0.712  \, \textless{}0.001                                                                                  & 0.613   \,  0.758  \, \textless{}0.001                                                                                                                                                                 \\ \hline 
VCG                              &      0.372 \, \textbf{1.019}  \, \quad $-$                                                                                                     &                  0.445 \, \textbf{1.154} \, \quad $-$                                                                                       &   0.474   \, \textbf{1.266}  \, \quad $-$                                                                                                      &     0.470   \, \textbf{1.361}  \, \quad $-$                                                                                                                                                                                                                                                                                                                                                                      \\ \hline
JRegNet                          & \textbf{0.562$^{\dagger}$}    0.841  \, \textless{}0.001                                                                                & \textbf{0.641$^{\dagger}$}    0.955  \, \textless{}0.001                                                                                 & \textbf{0.687$^{\dagger}$}    1.041  \, \textless{}0.001                                                                                 & \textbf{0.725$^{\dagger}$}   1.108  \, \textless{}0.001                                                                                                                                                        \\ \hline
\end{tabular}
\caption{The results of experiments for different CTRs (Settings B1, B, B2 and B3). ``${\dagger}$'' indicates that the revenue has a significant improvement over other methods in paired t-test at $p < 0.05$ level.}\label{tbl:666666}
\end{table*}

\begin{table*}[]
\begin{tabular}{cllllll}
\hline
\multirow{2}{*}{\textbf{Method}} & {\multirow{2}{*}{{\begin{tabular}[c]{@{}c@{}}\multicolumn{1}{c}{\textbf{D1: 4 $\times$ 5 $\times$ 2}}\\  rev \quad \, sw \quad \, rgt\end{tabular}}}} & {\multirow{2}{*}{{\begin{tabular}[c]{@{}c@{}}\multicolumn{1}{c}{\textbf{D: 4 $\times$ 5 $\times$ 3}}\\  rev \quad \, sw \quad \, rgt\end{tabular}}}} & {\multirow{2}{*}{{\begin{tabular}[c]{@{}c@{}}\multicolumn{1}{c}{\textbf{D2: 4 $\times$ 5 $\times$ 4}}\\  rev \quad \, sw \quad \, rgt\end{tabular}}}} & {\multirow{2}{*}{{\begin{tabular}[c]{@{}c@{}}\multicolumn{1}{c}{\textbf{D3: 4 $\times$ 5 $\times$ 5}}\\  rev \quad \, sw \quad \, rgt\end{tabular}}}} \\
                                 & \multicolumn{1}{c}{}                                                                                      & \multicolumn{1}{c}{}                                                                                      & \multicolumn{1}{c}{}                                                                                      & \multicolumn{1}{c}{}                                                                                      & \multicolumn{1}{c}{}                                                                                      & \multicolumn{1}{c}{}                                                                                      \\ \hline \\ [-10pt] \hline
RegretNet           &       0.363    \, 0.470 \, \textless{}0.001                                                                                                  &      0.370         \, 0.488 \,  \textless{}0.001                                                                                             &                                      0.372   \, 0.491  \,  \textless{}0.001                                                                 &    0.372     \, 0.494  \,  \textless{}0.001                                               
                                 
                                 \\  \hline
IRegNet                        & 0.537  \, 0.615  \, \textless{}0.001                                                                                 & 0.583   \, 0.690  \, \textless{}0.001                                                                                 & 0.603 \, 0.728  \, \textless{}0.001                                                                                  & 0.604   \,  0.732  \, \textless{}0.001                                                                                                                                                                 \\ \hline 
VCG                              &      0.579 \, \textbf{1.079}  \, \quad $-$                                                                                                     &                  0.602  \, \textbf{1.232}  \, \quad $-$                                                                                       &   0.599 \, \textbf{1.329}  \, \quad $-$                                                                                                      &     0.593   \, \textbf{1.354}  \, \quad $-$                                                                                                                                                                                                                                                                                                                                                           \\ \hline
JRegNet                          & \textbf{0.620$^{\dagger}$}     0.900  \, \textless{}0.001                                                                                & \textbf{0.676$^{\dagger}$}     1.005  \, \textless{}0.001                                                                                 & \textbf{0.693$^{\dagger}$}    1.062  \, \textless{}0.001                                                                                 & \textbf{0.704$^{\dagger}$}    1.076  \, \textless{}0.001                                                                                                                                                        \\ \hline
\end{tabular}
\caption{The results of experiments for different number of advertising slots (Settings D1, D, D2 and D3). ``${\dagger}$'' indicates that the revenue has a significant improvement over other methods in paired t-test at $p < 0.05$ level.}\label{tbl:333333}
\end{table*}

\end{document}